\documentclass[showpacs,twocolumn,superscriptaddress]{revtex4}

\usepackage{doi}% ----------
\usepackage{hyperref}
\hypersetup{
  colorlinks=true,        %
  linkcolor=blue,         %
  citecolor=cyan,         %
}

\usepackage{graphicx}
\usepackage{dcolumn}
\usepackage{bm}
\usepackage{color}

\usepackage{amsmath}
\usepackage{amssymb}


\begin{document}

\title{Six-dimensional Myers-Perry rotating black hole cannot be overspun}

\author{Sanjar Shaymatov}
\email{sanjar@astrin.uz}

\affiliation{Ulugh Beg Astronomical Institute, Astronomicheskaya 33, Tashkent 100052, Uzbekistan}
\affiliation{ National University
of Uzbekistan, Tashkent 100174, Uzbekistan}
\affiliation{Tashkent Institute of Irrigation and Agricultural Mechanization Engineers,\\ Kori Niyoziy 39, Tashkent 100000, Uzbekistan }

\author{Naresh Dadhich}
\email{nkd@iucaa.in}

\affiliation{Inter University Centre for Astronomy and
Astrophysics, Post Bag 4, Pune 411007, India }

\author{Bobomurat Ahmedov}
\email{ahmedov@astrin.uz}

\affiliation{Ulugh Beg Astronomical Institute, Astronomicheskaya
33, Tashkent 100052, Uzbekistan}
\affiliation{ National University
of Uzbekistan, Tashkent 100174, Uzbekistan}
\affiliation{Tashkent Institute of Irrigation and Agricultural Mechanization Engineers,\\ Kori Niyoziy 39, Tashkent 100000, Uzbekistan }

%

%\date{\today}
\begin{abstract}
{Though under nonlinear accretion, all black holes in four and higher dimensions obey the weak cosmic censorship conjecture (CCC); however, they generally violate it for linear test particle accretion with the exception of five-dimensional rotating black hole with a single rotation. In dimensions greater than five, there exists no extremal condition for black hole with single rotation and hence it can never be overspun. However, the extremal condition does exist for five-dimensional black hole with two rotations and then it could indeed be overspun under linear accretion. In this paper, we study the case of six-dimensional rotating black hole with two rotations and show that unlike the five-dimensional black hole it cannot be overspun under linear accretion. Though for nonlinear accretion, this result is anyway expected to hold good, yet we have verified it with an explicit calculation. Further, we would like to conjecture that so should be the case in all dimensions greater than six. Thus, the weak CCC may always be obeyed even at linear accretion process for rotating black hole in all dimensions greater than five.}

\end{abstract}
 \maketitle

\section{Introduction}
\label{introduction}

{Since cosmic censorship conjecture \cite{Penrose69} has remained unproven, testing its validity in various circumstances and consideration of physical processes that may lead to destruction of event horizon laying bare singularity have attracted attention of many works. For the latter case, a gedanken experiment is envisioned in which test particles of suitable parameters imploded on black hole so as to overspin or overcharge a black hole, and thereby destroying its horizon and creating a naked singularity. Naked singularity is however one of the most important unanswered questions in general relativity. Its  theoretical existence is important because it would mean that it is possible to observe collapse of an object to infinite density. The formation of naked singularity in gravitational collapse has long history beginning with Chirtodoulou (1986)~\cite{Christodoulou86}, and several others, for instance~\cite{Joshi93,Joshi15,Joshi00,Goswami06,Harada02,Stuchlik12a,
Vieira14,Stuchlik14,Giacomazzo-Rezzolla11}. Black holes have taken the center stage after the LIGO-VIRGO detection of gravitational waves produced by stellar mass black hole mergers \cite{Abbott16a,Abbott16b}. Gravitational wave is new and very potent tool to explore black hole properties which have so far remained unexplored.}

{The question, could an extremal Kerr-Newman charged and rotating black hole be converted into a naked singularity with $M^2 > a^2 +Q^2$ by throwing in test particles of suitable parameters was first addressed by Wald \cite{Wald74b} and it was shown that it cannot be done.
What happens is that particles with large charge or spin cannot be absorbed by an extremal Kerr-Newman black hole. On the other hand, a question of converting nonextremal black hole into extremal was addressed \cite{Dadhich97} and it was found that parameter window required for particle to hit the horizon pinches off as extremality is approached. Hence, no nonextremal black hole can be converted into an extremal one by test particle accretion. Then a new and novel setting was envisioned that though extremality cannot be attained, but it may however be jumped over in a discontinuous manner. That is transition to over extremality is not through extremality but through a discontinuous jump over extremality. This was initiated by Hubeny \cite{Hubeny99} in which a charged black hole was shown to be overcharged. Following that, Jacobson and Sotirio \cite{Jacobson10} had shown that a rotating black hole could be overspun by impinging black hole with test particles of suitable parameters. This opened up a new vista of investigations to study the phenomenon of destroying black hole horizon by overspinning/charging. In this thought, experiment one begins with a near extremal black hole and then let test particles of suitable parameters kick into the black hole, and thereby overextremalizing it. Of course, this is a linear order process in which backreaction and self-force are ignored. There is an extensive body of work considering various situations; here we give some representative references \cite[see,
e.g.][]{Jacobson10,Saa11,Hubeny99,Matsas07,Berti09,Shaymatov15,
Bouhmadi-Lopez10,Rocha14,Li13,Jana18,Duztas18,Duztas-Jamil18b,Duztas-Jamil18a}.}

Recently, the mechanism of destroying a Kerr-Newman-AdS black hole has been considered by neglecting radiative and self-force effects~\cite{Song18}. In all these works, it was assumed that test particle follows a geodesic or Lorentz force trajectory and backreaction effects {were neglected}. It has been argued that if self-force and backreactions effects are taken into account, particle that could overextremalize black hole may not indeed be able to reach the horizon~\cite{Barausse10,Zimmerman13,Rocha11,Isoyama11,Colleoni15a,Colleoni15b}. The question of inclusion of backreaction effects has also been studied for a regular black hole \cite{Li13} and magnetized Reissner-Nordstr\"{o}m black hole as well~\cite{Shaymatov19b}. Further, an extensive analysis involving complex scalar test fields around a black hole has also been considered in testing CCC~\cite[see, e.g.,][]{Semiz11,Toth12,Duztas13,Duztas14,Semiz15,Gwak19}. Recently, CCC has also been considered for a rotating anti-de Sitter (AdS) black hole~\cite{Gwak16,Natario16,Natario20}. Nonextremalization of black hole was also considered~\cite{Israel86,Mishra19} in the context of black hole dynamics~\cite{Bardeen73b}. It is worth
noting that the gedanken experiment~\cite{Wald74b} was also extended to an extremal magnetized black hole case~\cite{Siahaan16}. The
weak CCC was considered in the case of
BTZ black holes by throwing in test particles and
fields~\cite{Duztas16}. Chakraborty et al. \cite{Chakraborty17}
approached the issue by considering spin precession in the vicinity of black hole and naked singularity, showing a
clear distinction between the two.

This was all in the linear test particle accretion; however, Sorce  and Wald \cite{Sorce-Wald17,Wald18} have studied the gedanken experiment for nonlinear flow accretion, and have shown that black hole cannot be overextremalized and CCC is always obeyed. Following that An et al \cite{An18} have shown how the situation gets miraculously reversed when second order effects are taken into account. That is, black hole could be overextremalized at linear order accretion but not when second order effects are included. Similar conclusions were also obtained for nonlinear order effects~\cite{Gwak18a,Ge18,Ning19,Wang19,Yan-Li19,Jiang19}.

Very recently, an interesting feature of five-dimensional rotating black hole has been brought out \cite{Shaymatov19a} showing that though it could be overspun under linear accretion when both rotations are present but could not be when there is only one rotation. It is interesting that a five-dimensional black hole with single rotation defies the general result that CCC could be violated at the linear accretion. That is, a black hole with one rotation behaves radically differently from the one with two rotations. Then an interesting question arises, what happens in the case of a five-dimensional charged rotating black hole -- an analog of Kerr-Newman? The question is interesting because overcharging should be possible in this case while no overspinning. Strange may it sound but it is true that there exists no exact solution of Einstein-Maxwell equation describing a true analog of Kerr-Newman black hole in five dimensions. That is, it is not possible to put a charge on a five-dimensional rotating black hole \cite{Hawking99,Myers-Perry86}. The closest that comes to it is the minimally gauged supergravity charged rotating black hole \cite{Chong05}, and in that case it turns out that for single rotation the ultimate outcome for linear accretion depends on which of the particle parameter, charge, or rotation is dominant \cite{Shaymatov19c}. If angular momentum is larger than charge, black hole cannot be overextremalized while the opposite is the result if the case is other way round. For nonlinear accretion, however, it cannot be overextremalized and CCC always holds good.

In dimension $>5$, a black hole with one rotation has no extremal condition and hence the question of its overspinning does not arise. The natural question then arises is, what happens for a six-dimensional black hole with two rotations, could it be overspun or not for linear accretion. This is the question we wish to address in this paper and the answer turns out to be no; i.e., a six-dimensional rotating black hole cannot be overspun even for linear accretion process. We would like to conjecture that the same should be the case in all dimensions greater than six as well. Thus, CCC may always be obeyed for rotating black holes in all dimensions greater than and equal to six.

In Sec.~\ref{Sec:six}, we describe briefly the six-dimensional
rotating black hole which is followed by the main concern of the study --- whether overspinning is possible or not. We conclude with a discussion in Sec.~\ref{Sec:conclusion}.

\section{Six-dimensional rotating Myers-Perry black hole spacetime }\label{Sec:six}

Rotating Myers-Perry black hole metric \cite{Myers-Perry86}
in six dimensions in the Boyer-Lindquist coordinates is given by
\begin{eqnarray}\label{6D_metric}
ds^2&=&-\frac{\Delta}{\Sigma}\left(dt-a\sin^2\theta
d\phi-b\cos^2\theta
d\psi\right)^2+\frac{\Sigma}{\Delta}dr^{2}\nonumber\\
&+&\Sigma d\theta^2 +
\frac{\sin^2\theta}{\Sigma}\left[(r^2+a^2)d\phi-a
dt\right]^2\nonumber\\&+&\frac{\cos^2\theta}{\Sigma}\left[(r^2+b^2)d\psi-b
dt\right]^2 \nonumber\\&+&
r^2\left[\cos^2\theta+\sin^2\phi\right]\left(d\psi^2+\sin^2\psi d\eta^2\right)\ ,\ \quad
\end{eqnarray}
where
\begin{align}
&\Delta=\frac{\left(r^2+a^2\right)\left(r^2+b^2\right)}{r^2}-\frac{\mu}{r},
\label{eq:Delta}
\\
&\Sigma=r^{2}+a^{2}\cos^{2}\theta+b^{2}\sin^{2}\theta \, .
\label{eq:Sigma}
\end{align}
Here $\mu={8M}/{3\pi}$ is related to the black hole mass, while $a=3J_{\phi}/2M$ and $b=3J_{\psi}/2M$ are two rotation parameters of the black hole.
The event horizon is located at the largest real root of $\Delta=0$, and it is given by
\begin{widetext}
\begin{eqnarray}\label{eq:hor}
r_{\pm}&=&\frac{1}{2}\left[\frac{2^{1/3}(a^2+b^2)^2 7\pi}{\left(A+\sqrt{A^2-4B^3}\right)^{1/3}}-\frac{2}{3}(a^2+b^2)+\frac{\left(A+\sqrt{A^2-4B^3}\right)^{1/3}}{9\times 2^{1/3}\pi}\right]^{1/2} +\frac{1}{2}\left[\frac{16M}{3\pi}\left(\frac{2^{1/3}7(a^2+b^2)^2\pi}{\left(A+\sqrt{A^2-4B^3}\right)^{1/3}}\right.\right.\nonumber\\&-&\left.\left.\frac{2}{3}(a^2+b^2)+\frac{\left(A+\sqrt{A^2-4B^3}\right)^{1/3}}{9\times 2^{1/3}\pi}\right)^{-1/2}-\frac{2^{1/3}(a^2+b^2)^2 7\pi}{\left(A+\sqrt{A^2-4B^3}\right)^{1/3}}-\frac{4}{3}(a^2+b^2)-\frac{\left(A+\sqrt{A^2-4B^3}\right)^{1/3}}{9\times 2^{1/3}\pi}\right]^{1/2}\, ,
\end{eqnarray}
\end{widetext}
where
\begin{eqnarray}
A&=&54 \pi^3 \left(a^2+ b^2\right)^3-1944 \pi^3 a^2 b^2 \left(a^2+ b^2\right)\nonumber\\&+&5184 \pi  M^2 \, ,
\\
B&=& 108 \pi^2 a^2 b^2+9 \pi^2 \left(a^2+b^2\right)^2\, .
\end{eqnarray}
The extremality condition then reads $A^2-4B^3 = 0$; we write
\begin{eqnarray}\label{eq:extr}
A^2-4B^3 &=& 12\pi^2\left[768 M^4+16 \pi^2 M^2 \left(a^6-33 a^4 b^2\right.\right.\nonumber\\&- & \left.\left.33 a^2 b^4+b^6\right)-9 \pi ^4 a^2 b^2 \left(a^2-b^2\right)^4\right]\, .
\end{eqnarray}
Clearly, when one of rotations is zero, there occurs no extremal limit because the above expression is always positive. In general for $d>5$, there occurs no extremality limit for black hole with only one rotation parameter. For overspinning the above expression has to be negative.
%
%\begin{eqnarray}\label{eq:overspun}
%A^2-4B^3<0\, ,
%\end{eqnarray}
%
%is satisfied the black hole horizon can be overspun.

\subsection{Equal rotations, $a=b$}

The black hole horizon for $a=b$ is given by
\begin{eqnarray}
 r_{\pm}&=&6^{-1/2}A^{\prime 1/2}+\frac{1}{2}\left[8\sqrt{\frac{2}{3}}\pi^{-1}A^{\prime -1/2}-\frac{8a^2}{3}\right.\nonumber\\&-&\left.\frac{2^{7/3}a^4\pi^{2/3}}{3B^{\prime 1/3}}-\left(\frac{32}{27\pi^2}\right)^{2/3}B^{\prime 1/3}\right]^{1/2}\, ,
\end{eqnarray}
where
\begin{eqnarray}
&A^{\prime}=\left(\frac{4}{\pi^2}B^{\prime}\right)^{1/3}-2a^2+(16\pi^2)^{1/3}a^4B^{\prime -1/3}\, ,
\\
&B^{\prime}=3 M^2-2 \pi ^2 a^6+\sqrt{3}M \sqrt{3 M^2-4 \pi ^2 a^6 }\, 
\end{eqnarray}
and extremality is indicated by $3 M^2=4 \pi ^2 a^6$, which also follows from Eq.~(\ref{eq:extr}) when $a=b$. {However, if the following inequality:
\begin{eqnarray}\label{Eq:sin}
3 M^2<4 \pi ^2 a^6\, 
\end{eqnarray}
is satisfied, the final object turns into a naked singularity.
}
We consider the following three scenarios for test particle accretion:  (i) it has two equal rotations, $\delta J_{\phi}=\delta J_{\psi}=\delta J$; (ii) has only one rotation, $\delta J$ having projections on both rotation axes; and (iii) has only one rotation associated with $\delta J_{\psi}$.

(i) For equal rotation case, {from Eq.~(\ref{Eq:sin})}, the minimum threshold value would be defined by
\begin{eqnarray}\label{eq:min}
\sqrt[6]{\frac{3}{4\pi^2}} \bigg(M+\delta E\bigg)^{1/3}<\frac{3}{2}\left(\frac{J+\delta J}{M+\delta E}\right)\, .
\end{eqnarray}
That is, particle adds equal amount to both rotations of the black hole, $\delta J_{\phi}$ and $\delta J_{\psi}$. From the above equation, we write the minimum threshold value as
\begin{eqnarray}
\delta
J_{min}&=&\left(\frac{2}{3}\sqrt[6]{\frac{3}{4\pi^2}}{M}^{4/3}-J_{\psi}\right)+\frac{2}{3}\sqrt[6]{\frac{3}{4\pi^2}}~\frac{4}{3} M^{1/3}
\delta E\nonumber\\
&+&\frac{2}{3}\sqrt[6]{\frac{3}{4\pi^2}}~\frac{2}{9}M^{-2/3}\delta E^2\, .
\end{eqnarray}
We should here note that black hole is
endowed with two equal rotations and test particle also adds equal amount to both rotations of black hole about two axes.
We begin with a near extremal black hole, $a=b=\sqrt[6]{\frac{3}{4\pi^2}}M^{1/3}\left(1-\epsilon^2\right)$ with $\epsilon \ll 1$, and then $\delta J_{min}$ for either $\phi$ or $\psi$ rotation is given by

\begin{eqnarray}\label{Eq:min-threshold1}
\delta
J_{min}&=&\frac{2}{3}\sqrt[6]{\frac{3}{4\pi^2}}\left({M}^{4/3}\epsilon^2 +\frac{4}{3}~M^{1/3}
\delta E \right.\nonumber\\
&+&\left.\frac{2}{9}~M^{-2/3}\delta E^2\right)\, ,
\end{eqnarray}
and adding the two together, we write
\begin{eqnarray}
\left(\delta J_{\phi}+\delta J_{\psi}\right)_{min}&=&\frac{4}{3}\sqrt[6]{\frac{3}{4\pi^2}}\left({M}^{4/3}\epsilon^2 +\frac{4}{3}~M^{1/3}
\delta E \right.\nonumber\\
&+&\left.\frac{2}{9}~M^{-2/3}\delta E^2\right)\, .
\end{eqnarray}
This is the minimum threshold angular momentum for an impinging particle.

(ii) When infalling particle has only one rotation $\delta J$ having equal projections on both rotation axes of black hole. We will follow Eq.~(\ref{eq:extr}) to find the minimum threshold value of the particle angular momentum. In this case, we assume that a test particle's angular momentum $\delta J$ is equally shared between the two axes, $J_{\phi} +\delta J/2$ and $J_{\psi} +\delta J/2$. On accretion of particle, the black hole parameters change as $J_{\phi} +\delta J/2$, $J_{\psi} +\delta J/2$, and $M+\delta E$.

From Eq.~(\ref{eq:extr}), it then follows: %For equal rotation case $J_{\phi}=J_{\psi}$, the following must hold
\begin{equation}\label{naked} \frac{1024}{243\pi^2}\left({M+\delta E}\right)^2<
\left(\frac{2J_{\psi}+\delta J}{{M}+\delta E}\right)^{6}\, ,
\end{equation}
\\
leading to
\begin{eqnarray}
\delta
J_{min}&=&\left(\sqrt[6]{\frac{1024}{243\pi^2}}{M}^{4/3}-2J_{\psi}\right)+\frac{4}{3}\sqrt[6]{\frac{1024}{243\pi^2}} M^{1/3}
\delta E\nonumber\\
&+&\frac{2}{9}\sqrt[6]{\frac{1024}{243\pi^2}}M^{-2/3}\delta E^2.
\end{eqnarray}
Now writing $a=b=\sqrt[6]{\frac{3}{4\pi^2}}M^{1/3}\left(1-\epsilon^2\right)$, we obtain $\delta J_{min}$ as given by %Then $\delta J_{min}$ is given by
\begin{eqnarray}\label{jmin}
\delta J_{min} &=& \frac{4}{3}\sqrt[6]{\frac{3}{4\pi^2}}\left({M}^{4/3}\epsilon^2 +\frac{4}{3}~M^{1/3}
\delta E \right.\nonumber\\
&+&\left.\frac{2}{9}~M^{-2/3}\delta E^2\right)\, .
\end{eqnarray}
Note that this is the same as Eq.~(\ref{Eq:min-threshold1}) below indicating that the minimum threshold for angular momentum is the same in both the cases.

(iii) In the case of particle having a single rotation about one of the  black hole axes, the minimum threshold for angular momentum again turns out to be the same as given in the above equation. That is, in all three accretion scenarios, $\delta J_{min}$ is given by the same relation.

The maximum threshold value of the angular momentum allowed for particle to reach black hole horizon is defined by
\begin{eqnarray}\label{Eq:null}
\delta E \geq \Omega_{+}^{(\phi)}\delta J_{\phi}+\Omega_{+}^{(\psi)}\delta J_{\psi}\, ,
\end{eqnarray}
where $\Omega_{+}^{(\phi)}$ and $\Omega_{+}^{(\psi)}$ are angular velocities for two $\phi$ and $\psi$ axes, respectively. Thus, we have maximum threshold as
\begin{equation}\label{Eq:max} \delta
J_{max}=\delta J_{\phi}+\delta J_{\psi}=\frac{r^2_{+} + a^2}{a} \delta E \, ,
\end{equation}
which in view of {Eq.~(\ref{Eq:null})}, we write
{\begin{eqnarray}\label{Eq:max-threshold}
\delta J_{max}&=&
\left(1+\epsilon
+\frac{2}{3}~\epsilon^2\right)\frac{4}{3}\sqrt[6]{\frac{3}{4\pi^2}}~M^{1/3}~\delta
E\, .
\end{eqnarray}}

Hence, the difference between $\delta J_{max}$ and $\delta J_{min}$ for all cases (i--iii) takes the following form
{\begin{eqnarray} \label{eq:jmax-jmin2}
\Delta J &=& -\left[\left(\frac{1}{3}-\epsilon -\frac{2}{3}\epsilon^2 \right)~M^{1/3}~\delta E\right.\nonumber\\&+&\left. M^{4/3}\epsilon^2+ \frac{2}{9}~M^{-2/3}~\delta E^2\right]\frac{4}{3}\sqrt[6]{\frac{3}{4\pi^2}} \, ,
\end{eqnarray}}
which is clearly negative. Since $\Delta J < 0$ always, that indicates absence of parameter space available for overspinning and so horizon cannot be destroyed. Thus, CCC even under linear accretion is always respected. 

\subsection{Unequal rotations, $a \neq b$}

Let us here assume a particular case in which the ratio of two rotation parameters is $a=2b$. For this case, black hole horizon will have the form as 
\begin{eqnarray}
 r_{\pm}&=&\frac{\sqrt{3} }{6\pi^{1/3}}\left\{\left[\frac{73 \pi ^{4/3} b^4}{(A^{\prime\prime})^{1/3}}+(A^{\prime\prime})^{1/3}-10 \pi ^{2/3} b^2\right]^{1/2}\right.\nonumber\\&+&\left[\frac{16 \sqrt{3} M}{\sqrt{\frac{73 \pi ^{4/3} b^4}{(A^{\prime\prime})^{1/3}}+(A^{\prime\prime})^{1/3}-10 \pi ^{2/3} b^2}}-\frac{73 \pi ^{4/3} b^4}{(A^{\prime\prime})^{1/3}}\right.\nonumber\\&-& \left.\left.(A^{\prime\prime})^{1/3}-20 \pi ^{2/3} b^2\right]^{1/2}\right\} \, , 
\end{eqnarray}
with
\begin{eqnarray}\label{two-prime}
A^{\prime\prime}&=&96 M^2-595 \pi ^2 b^6\nonumber\\&+&4 \sqrt{576 M^4-2187 \pi ^4 b^{12}-7140 \pi ^2 b^6 M^2}\, .
\end{eqnarray}
This clearly indicates the extremality condition as
\begin{eqnarray}\label{Eq:sin1}
576 M^4 = 2187 \pi ^4 b^{12}+7140 \pi ^2 b^6 M^2\, .
\end{eqnarray}
From this, then follows the minimum threshold value required for overspinning, 
\begin{eqnarray}\label{eq:min}
\sqrt[6]{\frac{\alpha}{\pi^2}} \bigg(M+\delta E\bigg)^{1/3}<\frac{3}{2}\left(\frac{J+\delta J}{M+\delta E}\right)\, ,
\end{eqnarray}
where $\alpha$ is constant $\ll 1$. As mentioned in previous subsection, we have considered three scenarios for test particle accretion. However, in this case, we assume that particle adds angular momentum to the black hole in the ratio, $\delta J_{\phi}=2\delta J_{\psi}$. In doing so, we define the minimum threshold value as
\begin{eqnarray}
\delta
J_{min}&=&\left(\frac{2}{3}\sqrt[6]{\frac{\alpha}{\pi^2}}{M}^{4/3}-J_{\psi}\right)+\frac{2}{3}\sqrt[6]{\frac{\alpha}{\pi^2}}~\frac{4}{3} M^{1/3}
\delta E\nonumber\\
&+&\frac{2}{3}\sqrt[6]{\frac{\alpha}{\pi^2}}~\frac{2}{9}M^{-2/3}\delta E^2\, .
\end{eqnarray}
As usual we start out with a nearly extremal black hole, $b=\sqrt[6]{\frac{\alpha}{\pi^2}}M^{1/3}\left(1-\epsilon^2\right)$ with $\epsilon \ll 1$, and then $\delta J_{min}$ for $\psi$ rotation is given by
\begin{eqnarray}\label{Eq:min-threshold1}
\delta
J_{min}&=&\frac{2}{3}\sqrt[6]{\frac{\alpha}{\pi^2}}\left({M}^{4/3}\epsilon^2 +\frac{4}{3}~M^{1/3}
\delta E \right.\nonumber\\
&+&\left.\frac{2}{9}~M^{-2/3}\delta E^2\right)\, .
\end{eqnarray}
For both rotations ($\delta J_\phi = 2\delta J_\psi)$, the minimum threshold value is given by
\begin{eqnarray}
\left(\delta J_{\phi}+\delta J_{\psi}\right)_{min}&=&2\sqrt[6]{\frac{\alpha}{\pi^2}}\left({M}^{4/3}\epsilon^2 +\frac{4}{3}~M^{1/3}
\delta E \right.\nonumber\\
&+&\left.\frac{2}{9}~M^{-2/3}\delta E^2\right)\, .
\end{eqnarray}
Next, let us find the maximum threshold value of the angular momentum.  We recall Eq.~(\ref{Eq:null}) and rewrite the maximum threshold value as 
\begin{eqnarray}\label{Eq:max2} 
\delta J_{max}=\frac{\left(r^2_{+} + a^2\right)\left(r^2_{+} + b^2\right)}{2a\left(r^2_{+} + b^2\right)+b\left(r^2_{+} + a^2\right)} \delta E \, .
\end{eqnarray}
From the above equation, we write 
\begin{eqnarray}\label{Eq:max-threshold2}
\delta J_{max}&=&
2\left(1+\epsilon^2\right)\sqrt[6]{\frac{\alpha}{\pi^2}}~M^{1/3}~\delta
E\, .
\end{eqnarray}
This is the maximum threshold angular momentum. The difference between $\delta J_{max}$ and $\delta J_{min}$ is then given by 
\begin{eqnarray} \label{eq:jmax-jmin3}
\Delta J &=& -2\left[\left(\frac{1}{3}-\epsilon^2 \right)~M^{1/3}~\delta E\right.\nonumber\\&+&\left. M^{4/3}\epsilon^2+ \frac{2}{9}~M^{-2/3}~\delta E^2\right]\sqrt[6]{\frac{\alpha}{\pi^2}} \, .
\end{eqnarray}
The above expression is always negative definite. As always, $\Delta J < 0$ indicates that there is no parameter space available for an impinging particle to overspin black hole. Thus, Myers-Perry black hole in six dimensions always obeys the CCC for linear particle accretion.

\subsection{ Nonlinear order perturbation}

This is now well accepted from the work of Sorce and Wald \cite{Sorce-Wald17} that nonlinear accretion process can
only give the correct and true result whether black hole could be overextremized or not. The results
obtained under linear order process have all been overturned when nonlinear perturbations were included,
and in turn CCC was always obeyed. In this context, it is interesting that six-dimensional Myers-Perry rotating black hole cannot be overspun even under linear order accretion. Since it cannot be overspun at the linear accretion, there is no reason to expect that it would be otherwise when second order perturbations are included. However, for the sake of completeness, we shall perform the second order perturbations as well and show that the result continues to hold good, and CCC would be thus obeyed in general for a six-dimensional rotating black hole. 

We begin with a near extremal black hole and consider the general case of two rotations $a=kb$ where $k=0, 1$, respectively, indicate single and equal rotations. We define a function of the black hole parameters,
\begin{eqnarray}
f&=&576 M^4-\frac{2187 \left(33 k^2+33 k^4-k^6-1\right) \pi ^2 J^6}{16 M^4}\nonumber\\ &-&\frac{14348907 k^2 \left(k^2-1\right)^4  \pi ^4 J^{12}}{16384 M^{12}}\, .
\end{eqnarray}
Extremality is indicated by $f=0$ while overspinning by $f<0$. In the particular case of $k=0$ the above equation takes the form 
\begin{eqnarray}
f=576 M^4+\frac{2187 \pi ^2 J^6}{16 M^4}\, .
\end{eqnarray}
This clearly shows that $f>0$ always; thus, there exists no extremality condition and hence the question of its overspinning does not arise. Further, it is known that in dimensions greater than {five} a rotating black hole with a single rotation has no extremality condition. Though in five dimensions, it has extremality condition yet it cannot be overspun \cite{Shaymatov19a}. Thus, a black hole with single rotation cannot be overspun in all higher dimensions greater than four, and what we have shown here is that six-dimensional black hole with two rotations also cannot be overspun under linear accretion.

Following Sorce and Wald~\cite{Sorce-Wald17}, we turn to the one-parameter family of perturbation function $f(\lambda)$ including linear and nonlinear order perturbations, which are deviation from the initial very small value of $f$.  In doing so, we try to answer the question---could extremality be jumped over? We therefore write
\begin{eqnarray} \label{Eq:func1}
f(\lambda)&=& 576 M(\lambda)^4\nonumber\\&-&\frac{2187 \left(33 k^2+33 k^4-k^6-1\right) \pi ^2 J(\lambda)^6}{16 M(\lambda)^4}\nonumber\\ &-&\frac{14348907 k^2 \left(k^2-1\right)^4  \pi ^4 J(\lambda)^{12}}{16384 M(\lambda)^{12}} \, ,
\end{eqnarray}
where
\begin{eqnarray} \label{Eq:func2}
M(\lambda)&=& M+\lambda~\delta E\, ,\nonumber\\
J(\lambda)&=&J+\lambda~\delta J\, ,
\end{eqnarray}
and $\delta E$ and $\delta J$ refer to first order particle accretion. For near extremal black hole, we write $f(0)=576M^4\epsilon^2$ and first and second order perturbations are given by 
%
 %\begin{widetext}
 \begin{eqnarray}
f(\lambda)=576M^4\epsilon^2+f_1\lambda+f_2\lambda^2+O(\lambda^3, \lambda^2\epsilon,\lambda\epsilon^2,\epsilon^3)\, .
\end{eqnarray}
%\end{widetext}
If we are able to reach $f(\lambda)<0$, that would indicate overspinning and thereby creation of a naked singularity violating CCC. To explore that, we evaluate its values. From Eqs. (\ref{Eq:func1}) and (\ref{Eq:func2}), let us substitute $f_1$ and $f_2$ in the above equation for first and second order perturbations, and write 
 \begin{widetext}
 \begin{eqnarray}\label{Eq:liner-per}
f(\lambda)&=&576M^4\epsilon^2\nonumber\\&+&\frac{9437184 M^{16}+2239488 \left(33 k^2+33 k^4-k^6-1\right) \pi ^2 M^8 J^6 +43046721  k^2\left(k^2-1\right)^4 \pi ^4 J^{12}}{4096 M^{13}}\Big(\delta E-\Omega\delta J \Big)\lambda\nonumber\\&+&\frac{1}{2}\left[\frac{9437184 M^{16}+2239488 \left(33 k^2+33 k^4-k^6-1\right) \pi ^2 M^8 J^6 +43046721  k^2\left(k^2-1\right)^4 \pi ^4 J^{12}}{4096 M^{13}}\Big(\delta^2 E-\Omega\delta^2 J \Big)\right.\nonumber\\&+&\frac{28311552 M^{16}-11197440 \left(33 k^2+33 k^4-k^6-1\right)  \pi ^2 M^8 J^6-559607373 k^2 \left(k^2-1\right)^4\pi ^4 J^{12} }{4096 M^{14}}\delta M^2\nonumber\\ &+& \frac{26873856 \left(33 k^2+33 k^4-k^6-1\right) \pi^2 M^8 J^5+1033121304 k^2\left(k^2-1\right)^4 \pi^4 J^{11} }{4096 M^{13}}\delta M\delta J\nonumber\\ &-&\left. \frac{16796160 \left(33 k^2+33 k^4-k^6-1\right) \pi ^2 M^9 J^4 +473513931 k^2\left(k^2-1\right)^4  \pi^4 M J^{10} }{4096 M^{13}}\delta J^2\right ]\lambda^2+O(\lambda^3, \lambda^2\epsilon,\lambda\epsilon^2,\epsilon^3)\, .\nonumber\\
\end{eqnarray}
\end{widetext}
For an optimal choice of linear particle accretion, we take the minimum possible value of $\delta E$ absorbed by black hole as per the Sorce and Wald ~\cite{Sorce-Wald17} new gedanken experiment. The optimal choice of linear particle accretion satisfies          
\begin{eqnarray}
&\delta E & = \frac{6 J}{M} \nonumber\\ &\times &\left[\frac{9 J^2}{M^2}+\left( \frac{\sqrt{3}J}{M}+\sqrt{\frac{16 M^2}{3 \sqrt{3} \pi J}-\frac{12 J^2}{M^2}}\right)^2\right]^{-1}\delta J \nonumber\\
&+& O(\epsilon^2)\, ,
\end{eqnarray}
which is the minimal energy for the particle to fall into the black hole having equal rotations defined by $k=1$. However,  for general case, $k\neq 1$ $\delta E$ turns out to be very long and complicated expression for explicit display. We therefore resort to numerical evaluation of $f(\lambda)$. With this $\delta E$, we evaluate $f(\lambda)$ for linear order in $\lambda$ numerically.  Let us take $\delta J=J\epsilon^2$ for test particle approximation to hold good and setting $M=1$; we write $f(\lambda)$ in terms of $J$ as
\begin{widetext}
\begin{eqnarray}\label{Eq:linear-perturbation1}
f(\lambda)&=& 576 \epsilon^2+ \left[\frac{43046721 \pi ^4 J^{12} k^2 \left(k^2-1\right)^4}{4096}+\frac{2187}{4} \pi ^2 J^6 \left(33 k^2+33 k^4-k^6-1\right)+2304\right] \nonumber\\ &\times &\left[54 \left(\frac{1}{\frac{\left[3 \left(3\pi C_{2}\right)^{1/2}+C_{3}\right]^2}{32 \pi }+81 J^2}+\frac{k^2}{\frac{\left[3 \left(3\pi C_{2}\right)^{1/2}+C_{3}\right]^2}{32 \pi }+81 J^2 k^2}\right)J^2\epsilon^2  \right. \nonumber\\ &-& \frac{6561 \pi ^2 \left(6561 \pi ^2 J^6 k^2 \left(k^2-1\right)^4+512 \left(33 k^2+33 k^4-k^6-1\right)\right) J^6 \epsilon^2}{4096 \left(\frac{43046721 \pi ^4 J^{12} k^2 \left(k^2-1\right)^4}{4096}+\frac{2187}{4} \pi ^2 J^6 \left(33 k^2+33 k^4-k^6-1\right)+2304\right)}-\left(\frac{3 \left(3\pi C_{2}\right)^{1/2}+C_{3}}{\left(\frac{\left[3 \left(3\pi C_{2}\right)^{1/2}+C_{3}\right]^2}{32 \pi }+81 J^2\right)^2}\right.\nonumber\\&+&\left.\left.\frac{ \left(3 \left(3\pi C_{2}\right)^{1/2}+C_{3}\right)k^2}{\left(\frac{\left[3 \left(3\pi C_{2}\right)^{1/2}+C_{3}\right]^2}{32 \pi }+81 J^2 k^2\right)^2}\right) \left(\frac{2048 \sqrt{2} C_{3} C_{4}}{6561 \pi ^{7/2} C_{1}^4}+\frac{65536 \sqrt{\frac{2}{3}} \left(C_{1}^2-64 J^4 \left(k^4+14 k^2+1\right)\right)}{81 \pi ^2 C_{1}^4 C_{2}^{1/2}}\right)\frac{27 J^2 \epsilon^3}{\sqrt{2 \pi }}\right]\lambda \nonumber\\&+& \mathcal O(\lambda^2)\, ,
\end{eqnarray}
\end{widetext}
where  
\begin{widetext}
\begin{eqnarray}
C_{1}&=& \Big\{J^6 \left[3 \left(2187 \pi^2 J^6+5632\right) k^2+\left(16896-26244 \pi ^2 J^6\right) k^4+\left(39366 \pi ^2 J^6-512\right) k^6- 26244 \pi ^2 J^6 k^8\right. \cr &+&\left.6561 \pi ^2 J^6 k^{10}-512\right]\Big\}^{1/3}\, ,\\
%\end{eqnarray}
%\begin{eqnarray}
C_{2}&=& \frac{64 J^4 \left(k^4+14 k^2+1\right)}{C_{1}}+C_{1}-16 J^2 \left(k^2+1\right)\, , \\
C_{3}&=& \left[2048\left(\frac{6}{C_{2}}\right)^{1/2}-27 \pi  \left(C_{1}+32 J^2 \left(k^2+1\right)\right)-\frac{ 1728 \pi \left(k^4+14 k^2+1\right)J^4 }{C_{1}}\right]^{1/2}\, ,\\
%\end{eqnarray}
%\begin{eqnarray}
C_{4}&=& \frac{\frac{27 \pi  C_{1}^5}{J^6 \left(6561 \pi ^2 J^6 k^2 \left(k^2-1\right)^4-512 \left(k^6-33 k^4-33 k^2+1\right)\right)}+\frac{1024 \sqrt{6} \left(C_{1}^2-64 J^4 \left(k^4+14 k^2+1\right)\right)}{C_{2}^{3/2}}-1728 \pi  J^4 \left(k^4+14 k^2+1\right)}{\frac{6 J^4 \left(k^4+14 k^2+1\right)}{C_{1}}+\frac{3 C_{1}}{32}-\frac{64 \sqrt{\frac{2}{3}}}{3 \pi  C_{2}^{1/2}}+3 J^2 \left(k^2+1\right)}\, .
\end{eqnarray}
\end{widetext}

In particular, from Eq.~({\ref{Eq:func1}}), we write 
\begin{eqnarray}\label{Eq:num}
J&\text{=}&\frac{2 \sqrt{2}}{3 } \sqrt[3]{\frac{M^4}{3 \pi }}\left[\frac{1-33 k^2-33 k^4+k^6}{k^2 \left(k^2-1\right)^4}\right.\nonumber\\&+&\left.\frac{\sqrt{\left(k^4+14 k^2+1\right)^3-108 \epsilon^2 k^2 \left(k^2-1\right)^4}}{k^2 \left(k^2-1\right)^4}\right]^{1/6}\, \nonumber\\
\end{eqnarray}
for near extremality $f(0)=576M^4\epsilon^2$. However, for $k=1$, it takes the form 
\begin{eqnarray}\label{Eq:num1}
J\text{=}\sqrt[6]{\frac{16 M^8 \left(1-\epsilon^2\right)}{243 \pi ^2}}\, .
\end{eqnarray}
Then evaluating  Eqs.~(\ref{Eq:num}) and (\ref{Eq:linear-perturbation1}) numerically for different values of $k$ for given $J$, the results are as tabulated in Table~\ref{1tab}.
\begin{table}
\caption{\label{1tab} The value of $f(\lambda)$ for linear order in $\lambda $ for different values of $k$ and angular momentum $J$.  }
\begin{ruledtabular}
\begin{tabular}{c c c }
 $k$     & $J$  &     $f(\lambda)$    \\ \hline  
 1   & $0.433872$    &$\left[576+\left(1119.24+1.57\times 10^{-8}~\epsilon\right)\lambda\right]\epsilon^2$      \\[1ex]
 2   & $0.298021$    &$\left[576+\left(1144.92-0.568362~\epsilon\right)\lambda\right]\epsilon^2$      \\[1ex]
 3   & $0.233711$    &$\left[576+\left(1281.38-3.777150~\epsilon\right)\lambda\right]\epsilon^2$      \\[1ex]
 4   & $0.195215$    &$\left[576+\left(1526.11-11.173852~\epsilon\right)\lambda\right]\epsilon^2$      
\end{tabular}
\end{ruledtabular}
\end{table}
This clearly shows that  $f(\lambda)>0$ always for given $\lambda \lesssim \epsilon$. Thus, under linear order accretion, black hole cannot be overspun. 

Next, let us come to the second order quadratic in $\lambda$ perturbations for simplicity, $k=1$ corresponding to the black hole having equal rotations $a=b$, which is given by Eq.~(\ref{Eq:liner-per}). In linear order particle accretion, we considered null energy condition, $\delta E-\Omega \delta J = \int_{H} \Xi_{\alpha} \chi_{\beta} \delta T^{\alpha \beta}\geq 0$, where $\Xi_{\alpha}$ is volume element on the horizon. Thus, the first order perturbation refers to  the following inequality $\delta E\geq\Omega \delta J$. For the second order accretion, we take the variational inequality $\delta^2 E-\Omega \delta^2 J\geq -\frac{k}{8\pi}\delta^2 A$ \cite{Sorce-Wald17}. Here $\kappa$ and $A$ are surface gravity and area of the horizon, respectively, and  are given by
\begin{eqnarray}\label{Eq:SG}
\kappa &=&\frac{\Pi-\mu}{2\mu r_+}\, ,\\
\label{Eq:Hor-Area}
A&=&\frac{\Omega_{d-2}}{2k}\mu\left(d-3-2\frac{a^2(r_{+}^2+b^2)+b^2(r_{+}^2+a^2)}{\Pi}\right)\, ,\nonumber\\
\end{eqnarray}
with $\Omega_{d-2}=2\pi^{(d-1)/2}/\Gamma\left(\frac{d-1}{2}\right)$ and $\Pi=(r_{+}^2+a^2)(r_{+}^2+b^2)$. For $a=b$, we write for the second order perturbation, 
\begin{eqnarray}\label{Eq:second-order}
&&\delta^2 E-\Omega \delta^2 J \geq -\frac{\kappa}{8\pi}\delta^2 A=\nonumber\\
&&
2N\left(M,J\right)\left[ \left(54 \sqrt{3} \pi  J^5-8 J^2 M^4\right)\delta E^2 \right. \nonumber \\
&&+ \left(8 J M^5-72 \sqrt{3} \pi  J^4 M\right)\delta E\delta J \nonumber\\
&& +\left.\left(27 \sqrt{3} \pi  J^3 M^2-3 M^6\right)\delta J^2\right]\, ,
\end{eqnarray}
with
%
%\begin{widetext}
\begin{eqnarray}
N\left(M,J\right)&=&\frac{\left(4 \sqrt{3} M^4-27 \pi  J^3\right)^{-2} }{3 \sqrt{\pi } J^4 M \left(2 M \sqrt{\frac{4 \sqrt{3} M^4-27 \pi  J^3}{JM^2}}+3 \sqrt{3 \pi } J\right)}\nonumber\\ &\times & \left[32 M^8-243 \pi ^2 J^6-36 \sqrt{3} \pi  J^3 M^4\right. \nonumber\\ &+& \left(24 \sqrt{\pi } J^2 M^5 -54 \sqrt{3} \pi^{3/2} J^5 M \right)\nonumber\\ &\times & \left.\sqrt{\frac{4 \sqrt{3} M^4-27 \pi  J^3}{J M^2}}\right]\, .
\end{eqnarray}
%\end{widetext}
%
Further setting $M=1$ for second order perturbations, $f_{2}$ including terms for the second order quadratic in $\lambda$ then takes the following form:
\begin{widetext}
\begin{eqnarray}
f_{2} &\approx & 128\left[9\sqrt{3}\pi J^3\left(6J^2\delta E^2-8J\delta E \delta J+3\delta J^2\right)-\left(8 J^2 \delta E^2-8J \delta E \delta J+3\delta J^2\right)\right]\nonumber\\
&\times &\left[\frac{32-243 \pi ^2 J^6-36 \sqrt{3} \pi  J^3+24 \sqrt{\pi } J^2\sqrt{\frac{4 \sqrt{3}-27 \pi  J^3}{J}} -54 \sqrt{3}\pi^{3/2} J^5\sqrt{\frac{4 \sqrt{3}-27 \pi  J^3}{J}} }{3 \sqrt{\pi } J^4 \left(4 \sqrt{3}-27 \pi  J^3\right)^2 \left(2 \sqrt{\frac{4 \sqrt{3}}{J}-27 \pi  J^2}+3 \sqrt{3 \pi } J\right)}\right] +448\delta E^2-3645\pi^2 J^4\delta J^2\, .\nonumber\\
\end{eqnarray}
\end{widetext}
As before let us evaluate $f_2$ numerically. For given $\delta J=J\epsilon^2$ with $J=0.433872$, we have $f_2=651922~\epsilon^4$.  With this, it is clear that $f(\lambda)\geq 0$ always because  nonliner term $f_2$ is positive definite. Thus, nonlinear accretion always favors CCC.

Next, let us consider $k=2$ for second order quadratic in $\lambda$ perturbations,
%
%\begin{widetext}
 \begin{eqnarray}\label{Eq:non-linear-aneqb}
 &f_2&=\frac{9}{2048M^{13}}\left[\left(262144 M^{16}+37013760 \pi ^2 M^8 J^6 \right.\right.\nonumber\\&+& \left.387420489 \pi ^4 J^{12}\right)\left(\delta^2 E-\Omega\delta^2 J \right)\nonumber\\&+&\left(786432 M^{16}-185068800 \pi ^2 J^6 M^8\right.\nonumber\\&-& \left. 5036466357 \pi ^4 J^{12}\right) \delta M^2/M\nonumber\\ &+&  \left(444165120 \pi ^2 M^8 J^5 +9298091736 \pi ^4 J^{11}\right) \delta M \delta J \nonumber\\ &-&\left.  \left(277603200 \pi ^2 M^{9} J^4 +4261625379 \pi ^4 J^{10} M\right) \delta J^2\right]\nonumber\\ \, .
\end{eqnarray}
%\end{widetext}
As before we take $\delta^2 E-\Omega \delta^2 J\geq -\frac{k}{8\pi}\delta^2 A$ for second order perturbations. Then, recalling Eqs.~(\ref{Eq:SG}) and (\ref{Eq:Hor-Area}), we also evaluate $f_2$ numerically. For given $\delta J=J\epsilon^2$ with $J=0.298021$ and $M=1$, we have 
\begin{eqnarray}
f_2\approx 5.93961\times 10^8 \epsilon^4\lambda^2\, .
\end{eqnarray}
This clearly shows that $f_2\geq 0$ for given $\lambda \lesssim \epsilon$.  Thus, $f(\lambda)\geq 0$ always, thereby indicating that black hole is not able to corssover to over extremality. So, even for unequal rotations, there occurs no overspinning of black hole and CCC is always obeyed. Thus, the result of no overspinning obtained for linear order accretion, as expected, continues to hold good for nonlinear accretion as well.

\section{Conclusions}
\label{Sec:conclusion}

For linear accretion process, it was possible to overspin black hole in four dimensions. A subtler behavior ensues when one goes to five dimensions where overspinning is not possible when black hole has only one rotation but it is possible when it has two rotations \cite{Shaymatov19a}. It has however been verified for one as well as for two rotations that when second order perturbations are included, as expected there is no overspinning permitted in either case. A black hole with one rotation in five dimensions defies what is true for four-dimensional rotating black hole and behaves radically differently from the one with two rotations.

Another interesting aspect of rotating higher dimensional black hole is that it has no extremal limit defined for a black hole with one rotation in dimension $>5$, and hence a rotating black hole with one rotation in higher dimension $>4$ can never be overspun. A higher dimensional black hole with one rotation thus always obeys CCC.

Next question is, what happens in six dimensions for a black hole with two rotations? This is the question we have addressed here and have shown that it could not be overspun even under linear accretion process. As nonlinear accretion process always favors CCC, hence  inclusion of second order perturbations would further reinforce CCC. For the sake of completeness, we have shown so with explicit calculation. Thus, six-dimensional rotating black hole always obeys CCC. What happens in the next higher dimension seven where a black hole can have three rotations. We would like to conjecture that the two rotations case here should be similar to one rotation case in five dimensions; i.e., it cannot be overspun and three rotations case should be similar to two rotations case in six dimensions, and again there should be no overspinning. Thus a rotating black hole is always expected to obey CCC even at linear accretion in all dimensions greater than five.

What has been alluded above as a conjecture, it would be interesting to further analyze subtler behavior of rotating black hole in higher dimensions. That is what we intend to take up next.

\section*{ACKNOWLEDGMENTS}

B.A. and S.S. acknowledge the Faculty of Philosophy and Science, Silesian University in Opava, Czech Republic, Inter-University Centre for Astronomy and Astrophysics, Pune, India, Goethe University, Frankfurt am Main, Germany, and Nazarbayev University, Nur-Sultan, Kazakhstan for warm hospitality. {N.D. wishes to acknowledge visits to Albert Einstein Institute,
Golm, ICTP, Trieste and to Astronomical Institute, Tashkent supported by the Abdus Salam International Centre for Theoretical Physics, Trieste under the Grant
No. OEA-NT-01.} This research is supported in part by Projects No. VA-FA-F-2-008 and No. MRB-AN-2019-29 of the Uzbekistan Ministry for Innovative Development, by the Abdus Salam International Centre for Theoretical Physics under the Office of External Activities, and by the Erasmus + Exchange Grant between Silesian University in Opava and National University of Uzbekistan.

\bibliographystyle{apsrev4-1}  %% BibTeX style
\bibliography{gravreferences}

 \end{document}